# *Ab-initio* analysis of X – ray absorption spectra of a metalloprotein: structure and vibrations

G. Veronesi,[1,#] C. Degli Esposti Boschi,[2] L. Ferrari,[1] G. Venturoli,[3] F. Boscherini,[1,4,♦]
F.D. Vila,[5] and J.J. Rehr[5]

[1] *Department of Physics and CNISM, University of Bologna,
viale C. Berti Pichat 6/2, I-40127 Bologna, Italy*
[2] *CNR, c/o CNISM, Department of Physics, University of Bologna,
viale C. Berti Pichat 6/2, I-40127 Bologna, Italy*
[3] *Department of Biology and CNISM, University of Bologna, via Irnerio 42, I-40126 Bologna, Italy*
[4] *CNR – IOM – OGG, c/o ESRF, BP 220, F-38043, Grenoble Cedex, France*
[5] *Department of Physics, University of Washington, Seattle, Washington, 98195, USA*

We present a comparison between Fe K – edge X – ray absorption spectra of carbonmonoxy – myoglobin and its simulation based on density functional theory determination of the structure and vibrations and spectral simulation with multiple scattering theory. An excellent comparison is obtained for the main part of the molecular structure, without any structural fitting parameters. The geometry of the CO ligand is reliably determined using a synergic approach to data analysis. The methodology underlying this approach is expected to be especially useful in similar situations in which high-resolution data for structure and vibrations are available.

PACS numbers: 61.05.cj, 33.20.Tp, 87.15.ag, 87.64.kd

[#] Present address: European Synchrotron Radiation Facility, BP 220, F-38043 Grenoble (France)

[♦] Corresponding author. E-mail: federico.boscherini@unibo.it

X – ray Absorption Fine Structure (XAFS) is a powerful tool for a high resolution description of the local equilibrium structure and vibrations. Much progress has been recently made as a result of improvements in instrumentation, theoretical understanding and analysis tools. An essential aspect of modern XAFS data analysis is the simulation of spectra based on the curved – wave multiple scattering formalism.[1,2]

In parallel to the development of the XAFS experimental technique, *ab – initio* simulations of the atomic and vibrational structure, such as those based on Density Functional Theory[3] (DFT), have increasingly become a powerful and reliable tool. It is thus not surprising that there is great interest both in the comparison between the results of experimental and computational probes of atomic structure and vibrations and in the possible synergies that might be found between these tools. In this paper we show a comparison between Fe K – edge XAFS spectra of a metalloprotein (carbonmonoxy – myoglobin, MbCO, see Fig. 1) and its simulation based on the structure and vibrations determined using a DFT – based method. The case of MbCO is particularly interesting because of its biological relevance and because taking into account the vibrational damping of the many strong multiple scattering (MS) contributions reliably and correctly is the subject of current research. We will show that without the need of any structural fitting parameters, we obtain an excellent comparison between experiment and simulations as far as the contributions to the spectra of the heme plane and the proximal histidine are concerned. At the same time, we found the simulations to be essential to determine the geometry of the CO ligand, illustrating how notable improvements can be obtained with a synergic approach to data analysis.

It is well known that the XAFS signal can (in a simplified notation) be written as

$$\chi(k) = S_0^2 \sum_{j=paths} N_j f_j(k) \, \sin[kr_j + \varphi_j] e^{-2k^2\sigma_j^2} \tag{1}$$

where $k$ is the photoelectron wave number, $S_0^2$ is the many – body amplitude reduction factor, $N_j$ is the path degeneracy, $f_j$ and $\varphi_j$ are the effective scattering amplitude and phase shift, $r_j$ the path length and $\sigma_j^2$ is the Debye – Waller Factor (DWF) which takes into account thermal vibrations. In the simplest case of a single scattering (SS) contribution involving the absorbing atom "0" and atom $i$ and moderate disorder described by a Gaussian distribution, the DWF is

$$\sigma_i^2 = \left\langle [\hat{r}_{0i} \bullet (\vec{u}_0 - \vec{u}_i)]^2 \right\rangle \tag{2}$$

where $\langle \cdots \rangle$ denotes a configurational average, $\hat{r}_{0i}$ is the unit vector joining the two atoms and $\vec{u}_i$ is the instantaneous deviation of the position of atom $i$ from its average position. MS contributions are similarly damped. The importance of correctly accounting for the effect of vibrations on XAFS spectra has been long recognized. There are various reasons for this. Determination of DWFs can be difficult in the data analysis of spectra of complex structures exhibiting a distribution of interatomic distances; this can lead to ambiguities or even errors in the structural determination, a case often found in light atom molecular structures of biological relevance. In highly symmetric atomic structures values for the DWFs can sometimes be estimated semiempirically from the Debye or Einstein model. Moreover, much physical insight on the vibrational properties can be gained from a detailed study of DWFs, for example on thermal expansion and anisotropies of thermal motion.[4]

The interest in a reliable estimation and determination of DWFs in complex structures has been recognized by the scientific community. In an early paper, Loeffen and Pettifer[5] used the known structure and vibrational density of states of Zn tetraimidazole to compare XAFS experiment to simulation, including all relevant MS contributions and no free parameters; they found that the differences between the two were significant, ascribing this to the status of spectral simulation theory. Dimakis and Bunker have focused on the particular issue of metalloproteins and have proposed a parametrization of DWFs on the first shell distances and angles for Zn[6] and Fe[7] metalloproteins; their method is based on DFT calculations of the structure and vibrations for selected, isolated, metal containing reduced clusters (e.g. Zn – cysteine) from which the full local structure can be built. This method is certainly useful and has been used by some of the present authors to determine the previously unknown site of metal ions in metalloproteins.[8] However, since it is a building block approach, it inherently does not take into account interactions between the various molecular clusters. A general approach to calculate DWFs in an arbitrary structure with a known dynamical matrix has been proposed by some of the present authors.[9] This work also highlighted the extreme sensitivity of DWFs to the equilibrium structure.

The main function of myoglobin is to store oxygen in muscles. Its active site, an $Fe^{2+}$ ion at the center of the heme plane, binds a histidine residue (proximal His93) on one side of the plane and small ligands on the other; in MbCO a CO molecule is bound. A sketch of the local structure around $Fe^{2+}$, as obtained from the DFT simulations (see below) is reported in Fig. 1. MbCO has been studied in detail and can therefore be considered a test case for structural studies of active sites of metalloproteins. Previous relevant experimental structural determinations have been obtained by X – ray diffraction[10] and XAFS.[11,12,13] Chance et al.[12] have highlighted the difficulty of determining the position of the CO ligand from XAFS alone, due to the relative weakness of its contribution, a fact which is relevant in the context of the present results. The structure and vibrations of the heme – CO bond in MbCO and its interaction with surrounding molecular environment has been the subject of a detailed simulation using a combination of DFT and force field methods.[14] It was found that the heme structure is quite rigid while there is a strong interaction between the CO and a distal histidine residue (His64).

Horse – heart Mb was purchased from Sigma and dissolved (~ 10 mM) in a solution containing trehalose 200 mM and a phosphate buffer 20 mM at pH 7; the solution was equilibrated with CO and reduced by anaerobic addition of sodium – dithionite. Fe K – edge XAFS spectra were recorded in the fluorescence mode on the GILDA BM08 beamline of the European Synchrotron Radiation Facility. Spectra were recorded at 80 K with a total acquisition time of 50 s/point; continuous inspection of the near edge spectra during acquisition guaranteed the absence of radiation damage. Standard procedures were followed to extract the XAFS signal from the raw data, which is reported as the dotted curve in Fig. 2.

The DFT calculations for both the structural optimization and the dynamical matrix of ionic vibrations about the equilibrium positions were performed with plane-wave codes and pseudopotentials (of the Perdew-Burke-Ernzerhof type) shipped with the QuantumESPRESSO suite.[15] The starting coordinates for simulations were taken from X − ray diffraction.[10] We used a $20 \times 20 \times 15$ Å$^3$ supercell, slightly larger than that used in previous studies,[16] and an energy cut-off of 100 Ry for 100 Kohn-Sham states. The most demanding part of the calculation was the vibrational spectrum at the Gamma point, computed by means of density-functional perturbation theory.[17] The vibrational density of states is reported as an inset in Fig. 1.

From the dynamical matrix we calculated[9] DWFs for the highest amplitude contributions to the XAFS signal. In Table 1 we report the SS paths involving Fe obtained from the simulation and the corresponding DWFs. We note that DWFs estimated according to reference 7 are roughly 10% higher. In Tab. 2 we list the ten highest amplitude MS paths and their corresponding DWFs. Based on these results we simulated the XAFS signal using the curved wave MS code FEFF 8.2.[18] The scattering potential was obtained through Self Consistent Field calculations in a cluster of 4.4 Å radius around the absorber; the Hedin-Lundqvist form was chosen for its exchange-correlation term. All of the MS paths composed by up to 4 legs and with half-path length < 5 Å were summed using the Artemis code.[19]

The insightful comparison between experimental XAFS spectra and simulated ones in a molecular cluster such as this one, characterized by a high number of MS scattering paths with a dense distribution of path lengths, poses two specific problems. Firstly, the comparison rests upon a good choice for two non structural parameters which affect the spectrum: the many body amplitude reduction factor ($S_0^2$) and a shift of the energy origin ($\Delta E$). Secondly, a method must be found to identify the contribution of specific paths to the overall spectrum, especially if some discrepancies are present. Consequently, an original step – by – step procedure was adopted. In an initial step, only the first shell SS contributions ($R = 1 – 2$ Å) were considered, $S_0^2$ and $\Delta E$ were fitted[19] to best reproduce the experimental spectrum, all path lengths and DWFs being fixed to the values reported in Tab. 1; an excellent comparison (inset of Fig. 2) and a value $S_0^2 = 0.75$ were obtained. In the final step, $S_0^2$ was fixed to the value found previously, all SS paths and 200 MS paths were included; the path lengths and DWFs of SS paths and the ten highest amplitude paths reported in Tab. 2 were fixed to the DFT calculated values while a common DWF for the remaining ones was used as a fitting parameter, finding a value $\sigma_{MS}^2 = (6 \pm 4)\ 10^{-3}$ Å$^2$. The simulated spectrum thus obtained is reported as the continuous line in Fig. 2. It is clear that the overall comparison is quite good as far as the main oscillation frequencies and the amplitudes are concerned; however, a significant discrepancy around $k \sim 7$ Å$^{-1}$ is evident.

Inspection of the sum of paths relative to the CO molecule in the optimized geometry (CO-QE trace in Fig. 2) suggests that such contribution might be responsible for the discrepancy. In fact, in the DFT simulation the CO molecule was found to be essentially perpendicular to the heme plane (Fig. 1), giving rise to strong MS contributions due to the collinear geometry. However, there is ample evidence that the CO is actually tilted[11,12,13,14] when the heme plane and its axial ligands are embedded in their molecular environment (the protein's aminoacidic scaffolding), a fact which is impossible to take into account in a full DFT simulation at present due to limitations in computing power. To illustrate our reasoning we report in Fig. 2 the signals originating from CO in three geometries: as calculated by DFT, in the geometry determined from XRD[10] (labelled CO-1A6G) and according to Chance et al.[11] (labelled CO-Chance).

In the final step in our analysis we performed a new spectral simulation in which the coordinates of the heme plane and of His93 were fixed to the previously found values, while those of CO were as proposed by Chance et al.,[11] specifically, the Fe – C bond length was $R_{Fe-C}$ = 1.93(2) Å and the Fe – C – O bond angle $\theta_{Fe-C-O}$ = 127(4)°. Subsequently, a least – squares fit was performed in which $R_{Fe-C}$ and $\theta_{Fe-C-O}$ were treated as free parameters while, as previously, all path lengths and DWFs listed in Tab. 1 and 2 were kept fixed. It was found to be unnecessary to include other MS paths other that those listed in Tab. 2. Best fit values were $R_{Fe-C}$ = 1.78(2) Å and $\theta_{Fe-C-O}$ = 136(1)°, both characterized by a low uncertainty and significantly different both from the values proposed in reference 11 and also from results obtained by X-ray diffraction[10] ($R_{Fe-C}$ = 1.82(2) Å and $\theta_{Fe-C-O}$ = 171(3)°). The final comparison between the experimental spectrum and the simulation is very rewarding as can be seen from Fig. 3: virtually all spectral features are reproduced in frequency and amplitude. The $R$ – factor for this comparison is 7 %, a value which is similar to that obtained in XAFS fits with many free parameters in biomolecules.

We can make the following comments on the present result.

Firstly, we have shown that state – of – the – art simulation of the equilibrium structure and vibrations using DFT combined with the calculation of DWFs based on the dynamical matrix and XAFS spectral simulation compares very well with experiment in a complex molecular structure characterized by the presence of many scattering paths with a wide distribution of path lengths. This illustrates very favorably the reliability of present simulation methods and the calculation of DWFs[9].

Secondly, we have shown that use of simulations greatly enhances the robustness of XAFS data analysis. Thanks to the *ab – initio* determination of the structure and vibration of the heme plane and proximal histidine it was possible to selectively fit the contributions of the CO ligand to the overall spectrum, increasing the reliability of structural determination with respect to a full fitting, "brute force", approach.

Finally, we can conclude that in MbCO the structure and dynamics of the heme plane and proximal histidine are completely determined by internal interactions while the CO is sensitive to the molecular environment, in as much its position is not reproduced by simulations which do not take this effect into account. A related finding is that Gaussian disorder is sufficient to describe vibrational effects.

In conclusion, we believe these results indicate that in the future a closer interaction between experimental data analysis and *ab – initio* simulation is to be expected, in the field of XAFS but also of other high resolution probes of structure and vibration of condensed matter.

Acknowledgements

DFT calculations were performed at CINECA (Italy), "Iniziativa Calcolo per la Fisica della Materia" project number 898. Research supported in part by the European Theoretical Spectroscopy Facility. Gio. Vent. acknowledges the financial support of MIUR (Italy), grant PRIN 2008ZWHZJT. JJR acknowledges support from DOE Grant DE-FG03-97ER45623. We thank L. Cordone (University of Palermo) and F. Francia (University of Bologna) for useful discussions and help in sample preparation.

Table 1:

Single scattering paths involving $Fe^{2+}$ and corresponding DWF found from the DFT simulations.

| Path | Half path length (Å) | Degeneracy | $\sigma^2$ ($10^{-3}$ Å$^2$) |
|---|---|---|---|
| Fe-$N_{His}$ | 2.101 | 1 | 2.95 |
| Fe-$C_{CO}$ | 1.735 | 1 | 1.98 |
| Fe-$N_p$ | 2.012 | 4 | 2.41 |
| Fe-$C_2$ | 3.049 | 8 | 2.84 |
| Fe-$C_3$ | 3.428 | 4 | 2.97 |
| Fe-$C_4$ | 4.218 | 8 | 2.88 |

Table 2

Highest amplitude MS paths and corresponding DWFs. The top two paths are due to the CO ligand while the remaining ones to the heme plane.

| Path | Half path length (Å) | Degeneracy | $\sigma^2$ ($10^{-3}$ Å$^2$) |
|---|---|---|---|
| Fe-$C_{CO}$-$O_{CO}$-$C_{CO}$ | 2.897 | 1 | 1.97 |
| Fe-$C_{CO}$-$O_{CO}$ | 2.897 | 2 | 1.97 |
| Fe-$N_p$-$C_2$ | 3.220 | 16 | 2.68 |
| Fe-$C_4$-$N_p$ | 4.305 | 16 | 2.83 |
| Fe-$C_2$-$C_4$-$N_p$ | 4.411 | 16 | 2.82 |
| Fe-$N_p$-$C_2$-$N_p$ | 3.391 | 8 | 3.03 |
| Fe-$C_4$-$C_2$-$N_p$ | 4.558 | 16 | 3.19 |
| Fe-$C_4$-$C_2$ | 4.387 | 16 | 2.96 |
| Fe-$N_p$-$C_4$-$N_p$ | 4.329 | 8 | 2.87 |
| Fe-$C_2$-$C_3$ | 3.933 | 16 | 2.77 |

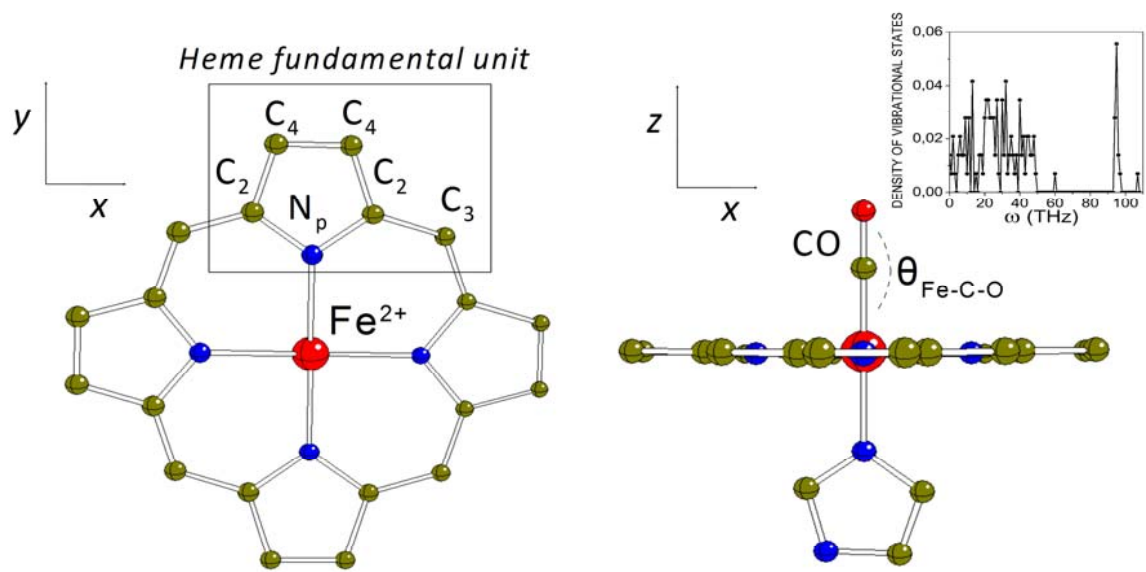

Fig. 1: (Color online) Graphical representation of the Fe site in MbCO. The inset reports the vibrational density of states calculated using density-functional perturbation theory.

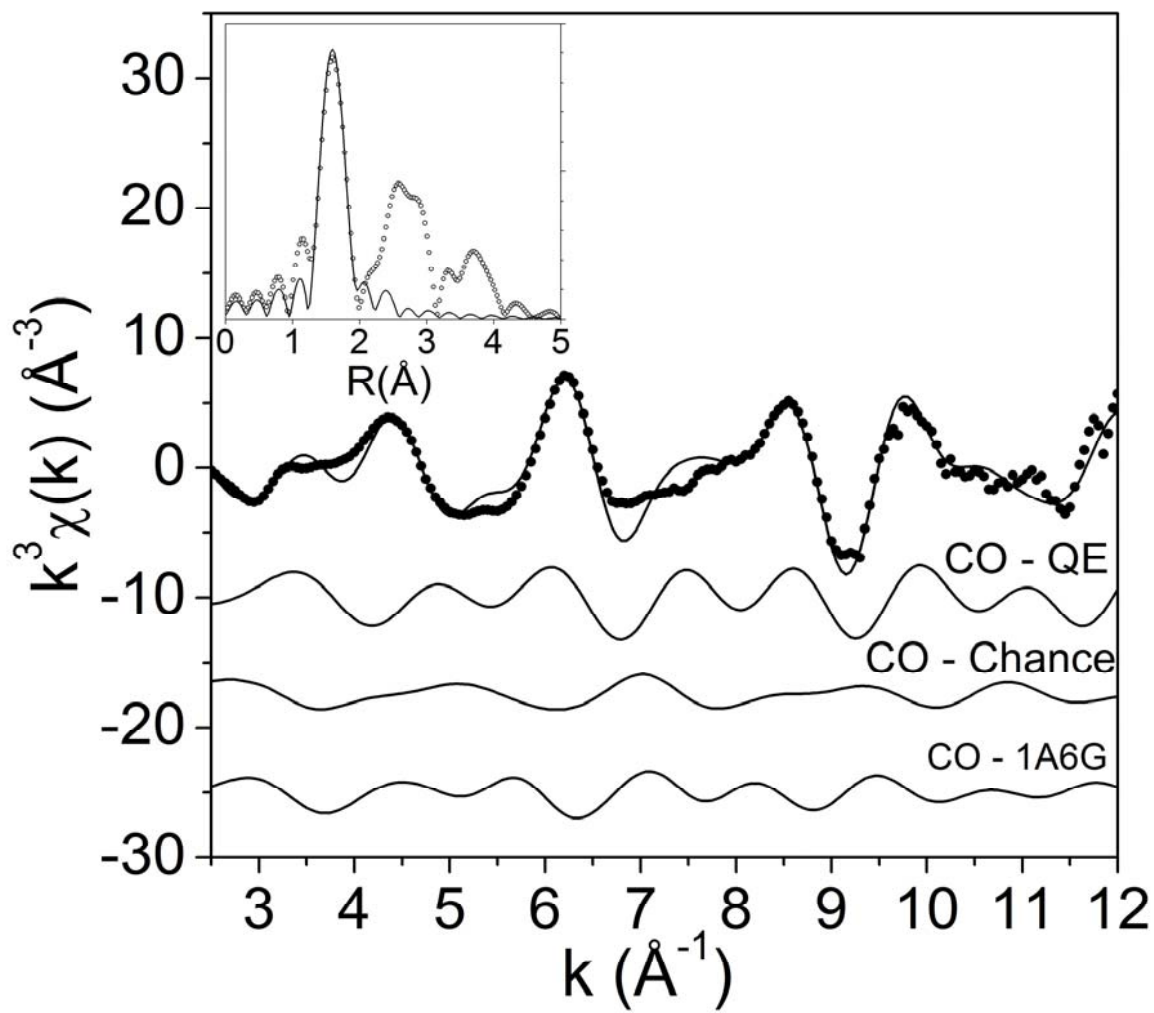

Fig. 2: Dots: experimental XAFS signal; continuous line: simulation based on the equilibrium DFT structure and estimation of DWFs. The bottom traces report the contributions of the CO ligand in various geometries (see text). In the inset we report the comparison of experiment and simulation for the first shell.

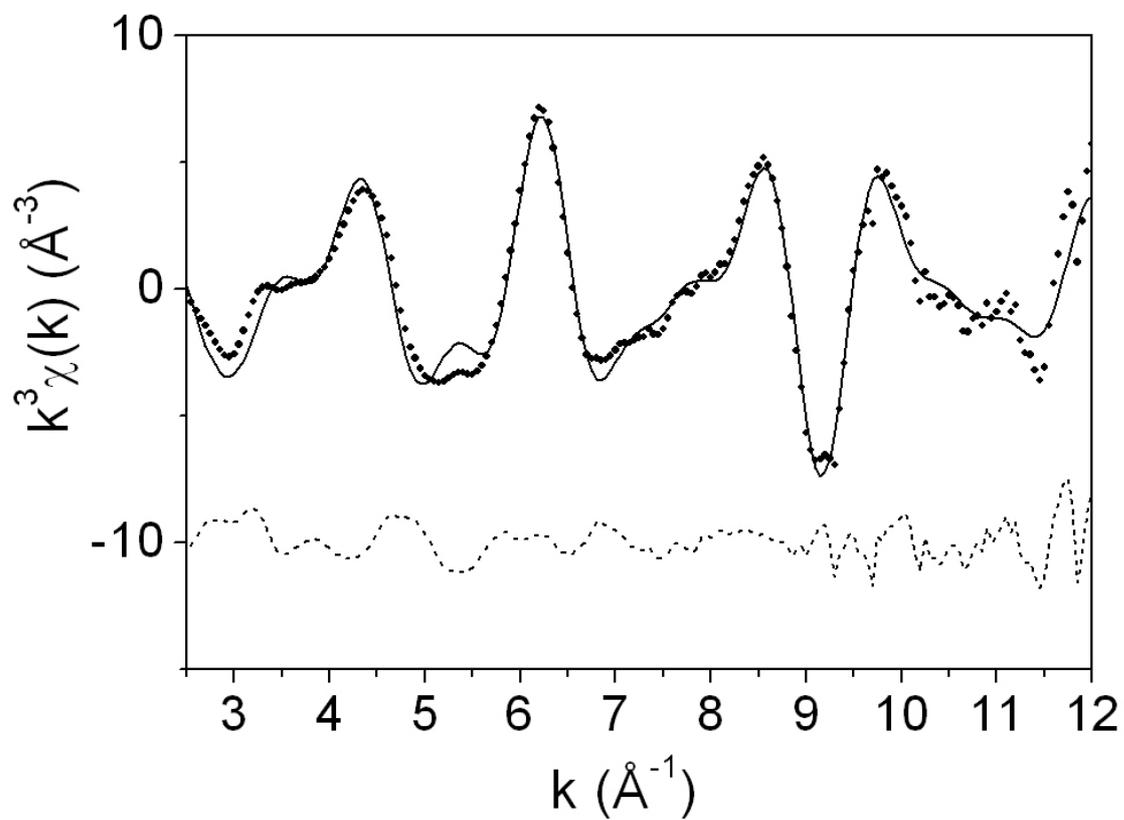

Fig. 3: Dots: experimental XAFS signal; continuous line: simulation based on DFT structure and vibration and fit of the geometry of the CO ligand, as explained in the text. The bottom trace (shifted for clarity) is the residual.